\documentclass[a4paper]{article}

\usepackage{amsfonts}

\def\be{\begin{equation}}
\def\ee{\end{equation}}
\def\bea{\begin{eqnarray}}
\def\eea{\end{eqnarray}}
\def\({\left(}
\def\){\right)}
\def\<{\left<}
\def\>{\right>}

\def\[{\left[}
\def\]{\right]}
\def\+{\bar}
\begin{document}

\pagestyle{empty}
\vskip-10pt
\vskip-10pt
\hfill 
\begin{center}
\vskip 3truecm
{\Large \bf
The non-Abelian tensor multiplet in loop space
}\\ 
\vskip 2truecm
{\large \bf
Andreas Gustavsson}\footnote{a.r.gustavsson@swipnet.se}\\
\vskip 1truecm
{\it Institute of Theoretical  Physics,
Chalmers University of Technology, \\
S-412 96 G\"{o}teborg, Sweden}\\
\end{center}
\vskip 2truecm
{\abstract{We introduce a non-Abelian tensor multiplet directly in the loop space associated with flat six-dimensional Minkowski space-time, and derive the supersymmetry variations for on-shell ${\cal{N}}=(2,0)$ supersymmetry.}}

\vfill 
\vskip4pt
\eject
\pagestyle{plain}
\section{Introduction}
The highest dimension in which one can have a superconformally invariant theory is $d=6$ \cite{Minwalla} and the maximally superconformal theory in $d=6$ has ${\cal{N}}=(2,0)$ chiral supersymmetry. The more symmetries we require on our theory, the better its quantum behaviour. One might hope that these maximally supersymmetic theories in six dimensions will enjoy the same finiteness property as their close relatives in four dimensions, ${\cal{N}}=4$ super Yang-Mills. Due to the difficulties with quantizing gravity, it has even been suggested that the $(2,0)$ theory might be the `theory of everything' \cite{Smilga}. According to that picture our universe would be a curved\footnote{In that way we get an induced gravity.} three brane embedded in flat six dimensions. Indeed the $(2,0)$ supersymmetry algebra allows for a central extension that involves a three brane (as well as a selfdual string) \cite{West}. Even though this picture might not comprise the whole truth, we think that $(2,0)$ theory is an interesting `intermediate' quantum theory which might be simpler to study than the full quantum theory of gravity, yet more complicated than Yang-Mills.

But it is problematic to quantize $(2,0)$ theory. The coupling constant is a fixed number $\sim 1$ due to self-duality and the dyonic charge quantization condition for strings in six dimensions. It may therefore not be possible to go from a classical theory to a quantum perturbation theory. It is possible that $(2,0)$ theory only exists as a quantum theory. But one way to obtain a related quantum theory would be if one could find solitonic solutions to some classical equations of motion. One should then be able to find a quantum theory by expanding quantum fields about this classical solution in a parameter which is related to the inverse tension of the extended object. 
    
In this Letter we will indeed derive the classical equations of motion, though in loop space. We will introduce a non-Abelian tensor multiplet in loop space which has to be subject to certain constraints. We then show that it closes the supersymmetry algebra on-shell, and thus get as a by product the non-Abelian equations of motion of the loop fields in the tensor multiplet. It thus appears to be the unique way in which to generalize the Abelian tensor multiplet.

\section{The tensor multiplet and its constraints in loop space}\label{constraints}
We will assume flat $d=1+5$ dimensional Minkowski space-time $M$ with metric tensor $\eta_{\mu\nu}=$diag$(-1,1,1,1,1,1)$ and Lorentz symmetry group $SO(1,5)$. The $(2,0)$-supersymmetry is generated by $16$ real supercharges transforming in the chiral representation $(4,4)$ of $SO(1,5)\times SO(5)$, where $SO(5)$ is an internal R-symmetry group. Our spinor conventions are the same as in \cite{Motl}, and these are collected in appendix. Requiring all this supersymmetry and no dynamical gravity, there is just one Abelian multiplet, namely the tensor multiplet. It consists of a two-form gauge potential $B_{\mu\nu}(x)$ with anti self-dual field strength $H_{\mu\nu\rho}(x)=-\frac{1}{6}\epsilon_{\mu\nu\rho\kappa\tau\sigma}H^{\kappa\tau\sigma}(x)$, five Lorentz scalars $\phi^A(x)$ (where $A$ is a vector index of $SO(5)$), and four real chiral (i.e. symplectic Majorana-Weyl) spinors $\psi(x)$ which transform in the same $(4,4)$-representation as the supercharges.

An Abelian two-form gauge potential $B_{\mu\nu}(x)$ in $M$ can alternatively be viewed in a parametrized loop space as a one-form,\footnote{In order to get clean equations, we only consider the integrated forms of the loop space fields, that is, instead of $A_{\mu s}$ we only consider the `zero mode' $A_{\mu}$ obtained by integration over $s$.}
\bea
A_{\mu}(C):=\int ds B_{\mu\nu}(C(s))\dot{C}^{\nu}(s).\label{repgauge}
\eea
Here $C$ denotes a parametrized loop $s\mapsto C^{\mu}(s)$ in $M$ and $s$ will always run over some fixed interval, say $s\in [0,2\pi]$. In \cite{Gustavsson} we also introduced Abelian loop fields corresponding to the other fields in the Abelian tensor multiplet, 
\bea
\phi_{\mu}^A(C)&:=&\int ds \dot{C}_{\mu}(s)\phi^A(C(s))\cr
\psi_{\mu}(C)&:=&\int ds \dot{C}_{\mu}(s)\psi(C(s)).\label{repferm}
\eea

We propose there is {\sl{some}} non-abelian generalization of these loop space fields. Apriori the non-abelian loop space fields like $A_{\mu s}(C)$ may depend in any non-local way on the loop $C$. There is also an interesting and potentially fruitful way of realizing them in terms of a local connection one-form and a local two-form, in such a way that the loop space fields themselves become local in a certain sense (see for instance \cite{Schreiber},\cite{Girelli}). There are problems in obtaining a non-abelian local action in terms of this local two-form -- that local theory becomes just a Maxwell theory \cite{Girelli}, thus a non-abelian generalization is not possible. But maybe a more subtle theory can be obtained for these local fields, which is a theory induced from a much simpler loop space theory that we will (in parts) obtain in this paper. I have also suggested another realization of these loop space fields in \cite{Gustavsson}. In this paper we will not make any assumptions on the `inner structure' of the loop space fields, apart from one very natural assumption, that they take values in the adjoint representation of a Lie algebra.

We introduce a derivative in loop space,
\bea
\partial_{\mu}(C):=\int ds\frac{\delta}{\delta C^{\mu}(s)}
\eea
and a gauge covariant derivative 
\bea
D_{\mu}(C):=\partial_{\mu}(C)+eA_{\mu}(C)
\eea
where $e$ is a coupling constant (that can not be determined by supersymmetry alone). In the sequel we will drop the arguments $C$. The gauge covariant field strength is $eF_{\mu\nu}=[D_{\mu},D_{\nu}]$. The gauge transformations act as (considering infinitesimal transformations generated by the loop field $\Lambda(C)$),
\bea
\delta A_{\mu}&=&\frac{1}{e}D_{\mu}\Lambda,\qquad \delta F_{\mu\nu}=[F_{\mu\nu},\Lambda]\cr
\delta \phi^A_{\mu}&=&[\phi^A_{\mu},\Lambda]\cr
\delta \psi_{\mu}&=&[\psi_{\mu},\Lambda]\label{gauge}
\eea

Let us first consider the Abelian case and then look for a natural non-Abelian generalization. The constraints on the Abelian loop fields are
\bea
\partial^{\mu}\phi_{\mu}^A&=&0\cr
\partial^{\mu}\psi_{\mu}&=&0
\eea
which is easily seen by computing 
\bea
\partial_{\nu}(C)\phi^A_{\mu}(C)=\int ds \dot{C}_{\mu}(s)\partial_{\nu}\phi^A(C(s))-\eta_{\mu\nu}\int ds \dot{C}^{\kappa}(s)\partial_{\kappa}\phi^A(C(s)).
\eea
We then see that $\partial^{\mu}\phi_{\mu}^A$ corresponds to a total derivative which vanishes when integrated over the loop.

How should these constraints be generalized to the non-Abelian case? The natural generalization should be to take the following gauge covariant non-Abelian constraints,\footnote{We could also add commutator terms and still get something gauge covariant. The ultimate check that these constraints are the right ones will not become apparent until we check the supersymmetry variations in the next section.}
\bea
D^{\mu}\phi_{\mu}^A&=&0\label{constr1a}\\
D^{\mu}\psi_{\mu}&=&0.\label{constr1b}
\eea
But now it is not consistent with supersymmetry to impose these constraints alone, without also imposing the constraint
\bea
[\phi^A_{\mu},\psi_{\nu}]=[\phi^A_{\nu},\psi_{\mu}].\label{constr2}
\eea
To see this, we impose the following supersymmetry variations of the Bose loop fields,
\bea
\delta \phi^A_{\mu}&=&-i\bar{\epsilon}\Gamma^A\psi_{\mu}\cr
\delta A_{\mu}&=&-i\bar{\epsilon}\Gamma_{\mu\kappa}\psi^{\kappa}
\eea 
and find that the supersymmetry variation of constraint becomes
\bea
\Gamma^A D^{\mu}\psi_{\mu} + \Gamma^{\mu\nu}[\psi_{\nu},\phi^A_{\mu}]=0.
\eea
Hence we see that supersymmetry implies that we must also impose the constraint (\ref{constr2}). Similarly, we should impose the constraint
\bea
[\phi^A_{[\mu},\phi^B_{\nu]}]=0.
\eea

Since the Fermi field $\psi_{\mu}$ thus is constrained, we introduce the somewhat simpler field
\bea
\psi:=\Gamma^{\mu}\psi_{\mu}
\eea
with no vector index, for which we find the relations
\bea
[\psi,\phi^A_{\nu}]=\Gamma^{\mu}[\psi_{\nu},\phi^A_{\mu}]
\eea
and
\bea
\Gamma^{\nu}[\psi,\phi^A_{\nu}]=[\psi^{\nu},\phi_{\nu}^A].\label{rel}
\eea

\section{${\cal{N}}=(2,0)$ supersymmetry}\label{susy}
\subsection{The fermions}
We now make the most general ansatz for the variation of the spinors compatible with Poincare invariance and dimensional analysis for the Fermi loop field, which is such that it reduces to the known Abelian transformation if we take the gauge group to be Abelian,
\bea
\delta_{\epsilon}\psi=\Big(\frac{1}{2}F_{\mu\nu}\Gamma^{\mu\nu}-D_{\mu}\phi_{\nu}^A\(\Gamma^{\mu\nu}+a\eta^{\mu\nu}\)\Gamma_A\cr
+\frac{1}{2}\[\phi^A_{\mu},\phi^B_{\nu}\]\(c\Gamma^{\mu\nu}\delta_{AB}+d\eta^{\mu\nu}\Gamma_{AB}\)\Big)\epsilon
\eea
Noting the constraints, we directly see that we can reduce this ansatz to just
\bea
\delta_{\epsilon}\psi=\(\frac{1}{2}F_{\mu\nu}\Gamma^{\mu\nu}-D_{\mu}\phi_{\nu}^A\Gamma^{\mu\nu}\Gamma_A+\frac{d}{2}\[\phi^A_{\mu},\phi^B_{\nu}\]\eta^{\mu\nu}\Gamma_{AB}\)\epsilon.
\eea
We begin with computing the commutor of two supersymmetry variations when acting on the Fermi loop field $\psi$, saving the Bose loop fields for later. Noting the variation of the field strength, 
\bea
\delta F_{\mu\nu}=2i\bar{\epsilon}\Gamma_{[\mu|\kappa|}D_{\nu]}\psi^{\kappa}
\eea
we then get
\bea
\[\delta_{\eta},\delta_{\epsilon}\]\psi&=&i\Gamma^{\mu\nu}\(\epsilon\bar{\eta}-\eta\bar{\epsilon}\)\Gamma_{\mu}D_{\nu}\psi\cr
&&+i\Gamma^{\mu\nu}\(\epsilon\bar{\eta}-\eta\bar{\epsilon}\)D_{\mu}\psi_{\nu}\cr
&&+i\Gamma^{\mu\nu}\Gamma_A\(\epsilon\bar{\eta}-\eta\bar{\epsilon}\)\Gamma^A D_{\mu}\psi_{\nu}\cr
&&+ie\Gamma^{\nu\mu}\Gamma_A\(\epsilon\bar{\eta}-\eta\bar{\epsilon}\)\Gamma_{\mu}\[\psi,\phi^A_{\nu}\]\cr
&&-id\Gamma_{AB}\(\epsilon\bar{\eta}-\eta\bar{\epsilon}\)\Gamma^A\[\psi^{\mu},\phi^B_{\mu}\].
\eea
To get here we have used constraint (\ref{constr2}). Then using a Fierz rearrangement and various gamma matrix identities (which we have collected in the appendix), we get
\bea
\begin{array}{lcrl}
\[\delta_{\eta},\delta_{\epsilon}\]\psi & = & -\frac{2i}{16}(\bar{\eta}\Gamma_{\eta}\epsilon) & \Big\{16D^{\eta}\psi\cr
&&&-\Gamma^{\eta}\(7\Gamma^{\mu}D_{\mu}\psi-(3e+4d)\Gamma^{\mu}\Gamma_A[\phi^A_{\mu},\psi]\)\cr
 & & & -8\(\Gamma^{\mu}D_{\mu} \psi^{\eta} - e \Gamma^{\mu}\Gamma_A [\phi^A_{\mu},\psi^{\eta}]\)\Big\}\cr
&&+\frac{2i}{16}(\bar{\eta}\Gamma_{\eta}\Gamma_C\epsilon) & \Big\{\Gamma^C\Gamma^{\eta}\(\Gamma^{\mu}D_{\mu}\psi-(3e-2d)\Gamma^{\mu}\Gamma_A[\phi^A_{\mu},\psi]\)\cr
&&&+8\Gamma^C\(\Gamma^{\mu}D_{\mu}\psi^{\eta}-e\Gamma^{\mu}\Gamma_A[\phi^A_{\mu},\psi^{\eta}]\)\cr
&&&+ 16e[\psi,\phi^{C\eta}]\Big\}\cr
&&-\frac{i}{192}(\bar{\eta}\Gamma_{\eta\omega\tau}\Gamma_{CD}\epsilon) & \Big\{-\Gamma^{CD}\Gamma^{\eta\omega\tau}\(\Gamma^{\mu}D_{\mu}\psi - e \Gamma^{\mu}\Gamma_A[\phi^A_{\mu},\psi]\)\cr
&&&+4(e-d)\delta_A^{[C}\Gamma^{D]}\Gamma^{\eta\omega\tau}\Gamma^{\nu}[\psi,\phi_{\nu}^A]\Big\}
\end{array}
\eea
For this to become a representation of the $(2,0)$-supersymmetry algebra, $\[\delta_{\eta},\delta_{\epsilon}\]=2i\(\bar{\epsilon}\Gamma^{\nu}\eta\)\partial_{\nu}$ (modulo a gauge transformation), we must take $d=e$ and the Fermi equations of motion to be
\bea
\Gamma^{\nu}\(D_{\nu}\psi^{\eta}-e\Gamma_A[\phi^A_{\nu},\psi^{\eta}]\) &=& 0\cr
\Gamma^{\mu}\(D_{\mu}\psi-e\Gamma_A[\phi^A_{\mu},\psi]\) &=& 0.
\eea
The second equation can be derived from the first by contraction with $\Gamma_{\eta}$ and using various constraints.\footnote{The steps needed to see this are
\bea
\Gamma_{\eta}\Gamma^{\mu}\Gamma_A[\phi^A_{\mu},\psi^{\eta}] &=& 2\Gamma_A[\phi^A_{\nu},\psi^{\nu}]+\Gamma^{\mu}\Gamma_A[\phi^A_{\mu},\psi] \cr
&=& 2\Gamma_A\Gamma^{\nu}[\phi^A_{\nu},\psi] + \Gamma^{\nu}\Gamma_A[\phi^A_{\nu},\psi]\cr
&=& -\Gamma^{\nu}\Gamma_A [\phi^A_{\nu},\psi].
\eea}
So we indeed have just one fermionic equation of motion, as one would expect.

\subsection{The bosons}
In order to compute two supersymmetry variations of the Bose fields it appears that we need to know how to vary $\psi_{\mu}$. To this end we define
\bea
\psi_{\mu s} := Q_{\mu\nu}(s)\Gamma^{\nu}\psi_s
\eea
where we have introduced the projector
\bea
Q_{\mu\nu}(s) := \frac{\dot{C}_{\mu}(s)\dot{C}_{\nu}(s)}{|\dot{C}(s)|^2}
\eea
where $|\bullet|^2$ denoted the Minkowskian length square. A short calculation using gamma matrix algebra (and $Q_{\mu}^{\mu} =1$) reveals that this definition implies that
\bea
\Gamma^{\mu}\psi_{\mu s} = \psi_s
\eea
If we also define the orthogonal projector
\bea
P_{\mu\nu} := \eta_{\mu\nu} - Q_{\mu\nu}
\eea
then we can write all the supersymmetry transformations entirely in terms of $\psi$:
\bea
\delta \phi^A_{\mu s} &=& -iQ_{\mu\nu}(s) \bar{\epsilon}\Gamma^A\Gamma^{\nu} \psi_s\cr
\delta A_{\mu s} &=& -iP_{\mu\nu}(s) \bar{\epsilon}\Gamma^{\nu}\psi_s
\eea
We now impose further constraints:
\bea
P_{\mu}{}^{\nu}(s)\phi^{A}_{\nu s} &=& 0\cr
Q_{\mu}{}^{\nu}(s)F_{\nu s,\rho t} &=& 0
\eea

Now it is easier to compute two supersymmetry variations of the bosons by first deriving the supersymmetry variation of $\psi_{\mu}$. Using the above definition, we find that
\bea
\delta \psi_{\mu} = \frac{1}{2}Q_{\mu\nu} \Gamma^{\nu\tau\rho}\epsilon F_{\tau\rho} + \Gamma^{\rho}\Gamma_A\epsilon D_{\rho} \phi^A_{\mu} + \frac{e}{2}\Gamma_{AB}\Gamma^{\rho}\epsilon [\phi^A_{\mu},\phi^B_{\rho}]
\eea
The notation here is such that $Q_{\mu\nu}F_{\tau\rho}$ means 
\bea
\int dt\int ds Q_{\mu\nu}(s)F_{\tau s,\rho t}
\eea
and $F_{\tau s,\rho t} = \frac{1}{e}[D_{\tau s},D_{\rho t}] = -F_{\rho t,\tau s}$, so that if we antisymmetrize in $\tau,\rho$ (as enforced by $\Gamma^{\tau\rho}$) then we get a result that is symmetric in $s,t$ and hence we get an unambigious result if $Q_{\mu\nu}$ is evaluated at $s$ or at $t$.

Then, upon using the above constraints, we find that
\bea
[\delta_{\eta},\delta_{\epsilon}]A_{\mu} &=& 2i\bar{\epsilon}\Gamma^{\eta}\eta F_{\eta\mu} + \frac{1}{e}D_{\mu} \Lambda\cr
[\delta_{\eta},\delta_{\epsilon}]\phi^A_{\mu} &=& 2i\bar{\epsilon}\Gamma^{\eta}\eta D_{\eta}\phi^A_{\mu} + [\phi^A_{\mu},\Lambda]
\eea
with gauge parameter
\bea
\Lambda &=& 2ie\(\bar{\epsilon}\Gamma_{\eta}\Gamma_C\eta\) \phi^{\eta,C}
\eea
which is the same gauge parameter as we found for the variation of the fermions,
\bea
[\delta_{\eta},\delta_{\epsilon}]\psi &=& 2i\bar{\epsilon}\Gamma^{\eta}\eta D_{\eta}\psi + [\psi,\Lambda].
\eea
Thus we could implement $(2,0)$-supersymmetry in a non-abelian field theory in loop space. It is true that we had to assume that our loop space fields live on a contraint surface in configuration space. These constraints are gauge invariant and supersymmetric. We are convinced that one can not relax anyone of these constraints. They are also natural because they go over to abelian constraints when taking abelian gauge group.

\section{Yang-Mills type of equations}
To get the Bose equations of motion, we make a supersymmetry variation of the Fermi equation of motion. We then find the Bianchi identity
\bea
D_{[\mu}F_{\nu\rho]}&=&0
\eea
and the Bose equations of motion
\bea
D^{\mu}F_{\mu\nu}+[\phi_A^{\mu},D_{\nu}\phi^A_{\mu}]+{\mbox{fermions}}&=&0\cr
D^{\mu}D_{\mu}\phi_{\nu}^A-\frac{1}{2}[\phi_{B,\nu},[\phi^B_{\mu},\phi^{A,\mu}]]+{\mbox{fermions}}&=&0.
\eea
To get these equations we have made use of all the constraints. 

One should notice the resemblance between equations we have presented and those of maximally supersymmetric Yang-Mills in five dimensions. 

We know how a rigorous dimensional reduction is carried out only in the abelian theory. Consider the abelian case and the scalar kinetic field energy term in the usual space-time action,
\bea
\frac{1}{8\pi}\int d^6 x (\partial_{\mu}\phi)^2.
\eea
The numberical value of $4\pi$ of the `abelian coupling constant' is required by selfduality \cite{Henningson}. We then compactify $x^5 \sim x^5 + 2\pi R$ and then we define five dimensional scalar field $\Phi$ as
\bea
\Phi &=& {2\pi R}\phi.
\eea
Then the action reduces to 
\bea
\frac{1}{8 \pi^2 R}\int d^5 x \frac{1}{2} (\partial_m \Phi)^2
\eea
where we have made the split $\mu = (m,5)$. Hence the gauge coupling of the reduced theory is 
\bea
g^2 &=& 8\pi^2 R
\eea
This implies that Kaluza-Klein modes in five dimensions have masses ($n$ is an integer)
\bea
M_{KK} = \frac{n}{R} = \frac{8\pi^2 n}{g^2}.
\eea
This is exactly the mass spectrum of instantons in four dimensional SYM, including the correct normalization. Hence instantons are KK modes coming from six dimensions \cite{Tong}.

It is not clear to us how a rigorous dimensional reduction should be performed on our loop space theory. It is clear though, that we can get SYM equations by taking six-dimensional space to be of the form ${\bf{R}}^{1,4}\times S^1$, and by restricitng to minimal loops that wind the circle of radius $R$ and then define the reduced Yang-Mills fields as
\bea
\Phi^A(x) &:=& \int ds \phi^A_{5s}(C_x)\cr
A_{m}(x) &:=& \int ds A_{m s}(C_x).
\eea
Here $C_x$ denotes the loop around the compact dimension, over the point $x\in {\bf{R}}^{1,4}$. 

Since we have not written down any action for our loop space fields, we do not know how to see how the five-dimensional coupling constant relates to the compactification radius in the loop space formalism, though clearly we should expect the relation be that $g^2 = 8\pi^2 R$ also in the non-abelian case.

\section{Selfduality and local theory}\label{localtheory}
It is well-known that a selfduality condition of the three-form field strength
\bea
H_{\mu\nu\rho}(x)=-\frac{1}{6}\epsilon_{\mu\nu\rho\kappa\tau\sigma}H^{\kappa\tau\sigma}(x),
\eea
characterizes $(2,0)$ theory and it leads to many subtleties, like the non-existence of a unique quantum theory on generic six-manifolds.

But in loop space we did never see any selfduality constraint. Could it just evaporate into nothing when we formulate the theory to loop space then? The answer I think lies in that as long as one does not specify what the loop space fields really are, there is no selfduality constraint in loop space. At least we could see that no selfduality constraint on $F_{\mu s,\nu t}$ needs to be imposed in order to realize supersymmetry (at least not for the zero modes) in loop space. If on the other hand one reduces the loop space theory to a local theory by inserting for instance a local representation of the loop space fields such as Eqs (\ref{repgauge}), then one will find that the supersymmetry variations that are induced on these local fields requires that $H$ is selfdual.

Explicitly we find that the supersymmetry variations of the local fields become
\bea
\delta \phi^A&=&-i\bar{\epsilon}\Gamma^A\psi\cr
\delta \psi&=&\(\frac{1}{12}H_{\kappa\tau\rho}\Gamma^{\kappa\tau\rho}+\partial_{\mu}\phi^A \Gamma^{\mu}\Gamma_A\)\epsilon\cr
\delta B_{\mu\nu}&=&-i\bar{\epsilon}\Gamma_{\mu\nu}\psi.
\eea
and the chirality condition of the supersymmetry parameter transmutes into a selfduality condition on $H$. The selfdual part of $H$ does not belong to the supermultiplet and can consistently be put equal to zero.

\section{Future work}
We have not included the non-zero modes in the analyzis (we only considered the zero modes, or the integrated loop space fields). We have also not specified the loop space fields very precisely. If it turns out that one can use the local representation of the loop space fields, then we should not need to have to think of how to include the so much trickier\footnote{The non-zero modes are not subject to any simple constraints like $\eta^{\mu\nu}D_{\mu s}\phi_{\nu t} = 0$ and it appears that more information is required about the inner structure of the loop space fields if we shall be able to properly include the non-zero modes in the supersymmetry variations.} non-zero modes of the loop space fields, in order to derive the induced supersymmetry variations of the local fields. This is work in progress.

\newpage
\appendix
\section{Spinor conventions}
We use the same conventions as \cite{Motl}, that is, we use eleven-dimensional gamma matrices $\Gamma^M$ and make the split $\Gamma^M=(\Gamma^{\mu},\Gamma^A)$ corresponding to the split $SO(1,10)\rightarrow SO(1,5)\times SO(5)$. We define 
\bea
\Gamma:=\Gamma^{012345}.
\eea
The (anti-)commutation relations between all these gamma matrices are
\bea
\{\Gamma^{\mu},\Gamma^{\nu}\}&=&2\eta^{\mu\nu}\cr
\{\Gamma^{\mu},\Gamma^A\}&=&0\cr
\{\Gamma^A,\Gamma^B\}&=&2\delta^{AB}\cr
\{\Gamma^{\mu},\Gamma\}&=&0\cr
[\Gamma^A,\Gamma]&=&0\label{acom}
\eea
We impose the following $SO(1,10)$-invariant Majorana condition on the spinors,
\bea
\bar{\psi}=\psi^T C.
\eea
Here $\bar{\psi}:=\psi^{\dag}\Gamma^0$ and the eleven-dimensional charge conjugation matrix $C$ has the properties
\bea
C^T&=&-C\cr
C^{\dag}C&=&1
\eea. 
Letting $V$ denote the linear space of such Majorana spinors, we then define the $SO(1,5)\times SO(5)$-invariant chiral subspaces
\bea
V_{\pm}:=\{\psi\in V:P_{\pm}\psi=\psi\}.\nonumber
\eea
where 
\bea
P_{\pm}:=\frac{1}{2}\(1\pm \Gamma\)
\eea 
As a consequence of (\ref{acom}), $\Gamma^{\mu}:V_{\pm}\rightarrow V_{\mp}$ and $\Gamma^A:V_{\pm}\rightarrow V_{\pm}$.

The gamma matrices have the properties
\bea
\(\Gamma^M\)^T&=&-C\Gamma^M C^{-1}\cr
\(\Gamma^{M_1\cdots M_p}\)^T&=&(-1)^{\frac{p(p+1)}{2}}C\Gamma^{M_1\cdots M_p}C^{-1}\cr
\Gamma^T&=&-C\Gamma C^{-1}
\eea
and 
\bea
\Gamma\Gamma_{\mu\nu\rho}=\frac{1}{6}\epsilon_{\mu\nu\rho\kappa\tau\sigma}\Gamma^{\kappa\tau\sigma}\nonumber
\eea
If $\epsilon,\eta\in V$ then we get
\bea
\bar{\eta}\Gamma_{M_1}\cdots\Gamma_{M_p}\epsilon=(-1)^p\bar{\epsilon}\Gamma_{M_p}\cdots \Gamma_{M_1}\eta
\eea

We will let SUSY parameters be $\epsilon_-,\eta_-,...\in V_-$. The spinor $\psi_+$ which is in the corresponding tensor multiplet will be of opposite chirality to that of the SUSY parameter, thus $\psi_+\in V_+$. 

We have that  
\bea
\bar{\eta}_-\Gamma^{\mu_1}\cdots\Gamma^{\mu_k}\epsilon_-&=&0 {\mbox{ if $k$ is even}}\cr
\bar{\eta}_-\Gamma^{\mu_1}\cdots\Gamma^{\mu_k}\psi_+&=&0 {\mbox{ if $k$ is odd}}\nonumber
\eea 
To see this we note that $\Gamma \epsilon=\epsilon \Leftrightarrow \bar{\epsilon}\Gamma=-\epsilon$. 

In eleven dimensions a complete set of matrices is
\bea
\left\{1,\Gamma^M,...,\Gamma^{M_1M_2M_3M_4M_5}\right\}
\eea
because $\Gamma^{012\cdots 10}=1$. The number of independent matrices is $2^{10}$ which is the number of components in a squarical matrix acing on the $2^5$-dimensional Dirac spinor representation. In six dimension we take as the complete set the matrices
\bea
\left\{\Gamma^{\mu_1\cdots \mu_k}\Gamma^{A_1\cdots A_l}\right\}
\eea
in such a way that 
\bea
\sum_{k,l}\({}^6_k\)\({}^5_l\)=2^{10}.
\eea
A particularly nice choice\footnote{We notice that $\({}^5_0\)+\({}^5_1\)+\({}^5_2\)=2^4$ and that $2^6.2^4=2^{10}$.} is to let $k=0,...,6$ and $l=0,1,2$.

Using the completeness and normalization properties\footnote{Completeness of a set of matrices $\Gamma^A$ means that any matrix $M$ may be expanded as $M=\sum C_A \Gamma^A$ and a normalization property $tr(\Gamma_A\Gamma^B)=\delta_A^B$ enable us to determine the coefficients $C_A=tr(M\Gamma_A)$.} of these matrices we may obtain the Fierz rearrangement for $\epsilon,\eta\in V_-$,
\bea
\epsilon\bar{\eta}-\eta\bar{\epsilon}&=&\frac{1}{16}\(-(\bar{\eta}\Gamma_{\eta}\epsilon)\Gamma^{\eta}+(\bar{\eta}\Gamma_{\eta}\Gamma_A\epsilon)\Gamma^{\eta}\Gamma^A\)\(1+\Gamma\)\cr
&&-\frac{1}{192}(\bar{\eta}\Gamma_{\mu\nu\rho}\Gamma_{AB}\epsilon)\Gamma^{\mu\nu\rho}\Gamma^{AB}
\eea
Here are some gamma matrix identities we have used in this paper,
\bea
\Gamma^{\mu\nu}\Gamma^{\eta}\Gamma_{\mu}&=&8\eta^{\nu\eta}-3\Gamma^{\eta}\Gamma^{\nu}\cr
\Gamma^{\mu\nu}\Gamma^{\eta}&=&\Gamma^{\eta}\Gamma^{\mu\nu}-4\eta^{\eta[\mu}\Gamma^{\nu]}\cr
\Gamma^{\nu\mu}\Gamma_{\eta\omega\tau}\Gamma_{\mu}&=&\Gamma_{\eta\omega\tau}\Gamma^{\nu}\cr
\Gamma_{AB}\Gamma^{CD}\Gamma^A&=&-4\delta_B^{[C}\Gamma^{D]}\cr
\Gamma_A\Gamma^{CD}\Gamma^A&=&\Gamma^{CD}
\eea

\end{document}